\documentclass[]{interact}

\usepackage{epstopdf}
\usepackage[caption=false]{subfig}

\usepackage[numbers,sort&compress]{natbib}
\bibpunct[, ]{[}{]}{,}{n}{,}{,}
\renewcommand\bibfont{\fontsize{10}{12}\selectfont}
\makeatletter
\def\NAT@def@citea{\def\@citea{\NAT@separator}}
\makeatother

\theoremstyle{plain}

\theoremstyle{definition}

\theoremstyle{remark}

\usepackage[nolists,tablesfirst]{endfloat}
\usepackage{amsmath,amsthm,amssymb,bm}
\usepackage{multirow}
\usepackage{algorithm,algorithmic}

\newcommand{\matr}[1]{\bm{#1}}

\begin{document}


\title{Approximately Optimal Subset Selection for Statistical Design and Modelling}


\author{
\name{Y. Wang$^\ast$\thanks{$^\ast$The author is now at University of Michigan, Ann Arbor, USA}, N.~D. Le, J.~V. Zidek\thanks{CONTACT Y. Wang. Email: wayneyw@umich.edu}}
\affil{University of British Columbia, Vancouver, Canada}
}

\maketitle

\begin{abstract}
We study the problem of optimal subset selection from a set of correlated random variables. In particular, we consider the associated combinatorial optimization problem of maximizing the determinant of a symmetric positive definite matrix that characterizes the chosen subset. This problem arises in many domains, such as experimental designs, regression modelling, and environmental statistics. We establish an efficient polynomial-time algorithm using the determinantal point process to approximate the optimal solution to the problem. We demonstrate the advantages of our methods by presenting computational results for both synthetic and real data sets. 
\end{abstract}

\begin{keywords}
Combinatorial optimization; Determinantal point processes; $D$-optimal design; Maximum entropy sampling; Genetic algorithms; Parallelizable search algorithm
\end{keywords}

\section{Introduction}\label{intro}	

This paper addresses the problem of making inferences about a set of random variables given observations of just a subset of them. This problem relates closely to the topic of design of experiments in classical statistics.  There an experimenter selects and runs a well planned set of experiments to optimize a process or system from well supported conclusions about the  behaviour of that process of system. In environmental statistics, for example, the experiment yields observations of a certain environmental process (temperature, air pollution, rainfall, etc) taken from a set of monitoring stations. Since usually maintaining all stations would be costly and hence infeasible, one may need to select only a subset of them. Another example is seen in variable selection in regression models.  There the task consists of finding a small subset of the available independent variables that does a good job of predicting the dependent variable. 

In either case, a well--defined optimality criteria is needed for evaluating designs. Formally, consider a set $N$ of $n$ points, called the design space, and a design size $k$, such that $k \leq n$. Our goal is to select a subset $K$ of $k$ points, such that observations taken at these points are maximally informative. Information can be measured by entropy, for example, and our goal is then to choose a subset that minimizes the resulting entropy, i.e., maximizes the amount by which uncertainty will be reduced by the information provided by the experiment. Consider a symmetric positive definite $n \times n$ matrix $\matr{L}$ indexed by the set $N$, for instance, a covariance matrix. Then the entropy associated with any $k$--element subset $K$ of $N$, up to a known positive affine transformation, is the logarithm of the determinant of the $k \times k$ principal submatrix $\matr{L}[K]$ with row and column indices in $K$ (see Caselton and Zidek \cite{CaseltonZidek} for details). The criterion coincides with what is called the $D$--optimal design. The problem now is to find a design among the set of all feasible designs that maximizes det($\matr{L}[K]$). In classical regression models, the optimization criteria are generally related to the notion of the (Fisher) information matrix. In this context, the $D$--optimal design objective is to maximize the determinant of the corresponding information matrix.

As demonstrated in Ko et al. \cite{KO1995}, this optimization problem is NP--hard.  Thus we propose a new approximation strategy to this combinatorial optimization problem based on the determinantal point process (DPP). This novel approximation algorithm is stochastic, unlike other existing methods in the literature, and always approaches the optimum as the number of iterations increases. The proposed algorithm can easily be parallelized; thus multiple computer processing units could be used simultaneously to increase computing power. As shown in Section~\ref{dpp}, our algorithm is computationally efficient as measured by its running time. 

The remainder of the paper is organized as follows. In Section \ref{overview}, we formally define the problem and give an overview of existing algorithms for finding/approximating the optimal solution. In Section \ref{dpp}, we introduce the DPP  and describe a solution approach based on it. Numerical results are given in Section \ref{examples} for a comparison of accuracy and efficiency of different approaches. We conclude with recommendations on the use of our algorithm in practice and comments on future research directions.

\section{Overview of algorithms for finding the optimal solution}\label{overview}

\subsection{Definitions and notation}
Let $N = \{1, 2, 3, ..., n\}$ where $n$ is a positive integer. We use $\matr{K}$ to denote a real symmetric positive definite matrix indexed by elements in $N$. Further, let $S$ be an $s$--element subset of $N$ with $1 \leq s \leq n$. Let $\matr{K}[S, S]$ denote the principal submatrix of $\matr{K}$ having rows and columns indexed by elements in $S$--note that $\matr{K}[S, S] = \matr{K}[S]$. Write $v_{N} (S) = \text{det}(\matr{K}[S])$ to denote the determinant of the matrix $\matr{K}[S]$. Our optimization problem is to determine
\begin{equation}\label{P}
	\max_{{S}: |S|=s, S \subset N} v_{N} (S),
\end{equation} and the associated maximizer $S$. 

\subsection{Finding a solution}
Numerous algorithms have been developed for solving/approximating the optimization problem, including both exact methods and heuristics. For small, tractable problems, Le and Zidek \cite{LeZidek} describe an algorithm based on complete enumeration that been implemented in the \texttt{EnviroStat v0.4-0 R} package \cite{Le2014}. Ko et al. \cite{KO1995} first introduced a branch--and--bound algorithm that guarantees an optimal solution. Specifically, the authors established a spectral upper bound for the optimum value and incorporated it in a branch--and--bound algorithm for the exact solution of the problem. Although there have been several further improvements, mostly based on incorporating different bounding methods \cite{Anstreicher1996,Anstreicher1999,Hoffman2001,Lee2000,LeeWilliams}, the algorithm still suffers from scalability challenges and can handle problem of size only up to about $n = 75$ \cite{Lee2001}.    

\subsubsection{Greedy algorithm} 
For large intractable problems heuristics, all lacking some degree of generality and theoretical guarantees on achieving proximity to the optimum, can be used to find reasonably good solutions. One of the best known is the DETMAX algorithm of \cite{Mitchella,Mitchellb}, based on the idea of exchanges, which is widely used by statisticians for finding approximate $D$--optimal designs. Due to the lack of readily available alternatives, Guttorp et al. \cite{Guttorp} use a greedy approach, which is summarized in Algorithm~\ref{alg:GREEDY}. Ko et al. \cite{KO1995} experiment with a backward version of the Algorithm~\ref{alg:GREEDY}: start with $S = N$, then, for $j = 1, 2, ..., n-s$, choose $l \in S$ so as to maximize $v_{N} (S \setminus \{l\})$, and then remove $l$ from $S$. They also describe an exchange method, which begins from the output set $S$ of the greedy algorithm, and while possible, choose $k \in N \setminus S$ and $l \in S$ so that $v_{N} (S \cup \{k\} \setminus \{l\}) > v_{N} (S)$, and replace $S$ with $S \cup \{k\} \setminus \{l\}$.

\begin{algorithm}[tbh]
\caption{Greedy Algorithm}
\label{alg:GREEDY}
\begin{algorithmic}
\REQUIRE Size $k$ and an empty set $S=\emptyset$.
\FOR{$i = 1, \dots, k$}
        \STATE Choose $s \in N \setminus S$ so as to maximize $v_{N} (S \cup \{s\})$.
        \STATE Set $S=S \cup \{s\}$.
\ENDFOR
\ENSURE Set $S$ with $k$ elements.
\end{algorithmic}
\end{algorithm}

\subsubsection{Genetic algorithm}
More recently, Ruiz--C\'{a}rdenas et al.\cite{Schmidt} propose a stochastic search procedure based on Genetic Algorithm (GA) \cite{Holland} for finding approximate optimal designs for environmental monitoring networks. They test the algorithm on a set of simulated datasets of different sizes, as well as on a real application involving the redesign of a large--scale environmental monitoring network. In general, the GA seek to improve a population of possible solutions using principles of genetic evolution such as natural selection, crossover, and mutation. The GA considered here consists of general steps described in Algorithm \ref{alg:GA}. The GA has been known to work well for optimizing hard, black--box functions with potentially many local optima, although its solution is fairly sensitive to the tuning parameters~\cite{Goldberg,Whitley}.

\begin{algorithm}[tbh]
\caption{Genetic Algorithm}
\label{alg:GA}
\begin{algorithmic}[1]
\STATE Choose at random an initial population of size $N_0$, that is, a set of $N_0$ possible solutions $S_1,...,S_{N_0}$.
\STATE Compute the fitness, that is, the value of the objective function $v_{N}(S_i)$, $i = 1,...,N_0$, for each of the solutions in the population.
\STATE \textit{Crossover}: choose a proportion, $p_{\text{cross}}$, of solutions from the population. These solutions are selected according to a fitness-dependent selection scheme. Among these selected solutions, pairs of solutions are formed at random.
\STATE \textit{Mutation}: choose a proportion, $p_{\text{mutprop}}$, of solutions from the population with equal probability. For each selected solution, each gauged site may be swapped, according to a mutation probability $p_{\text{mut}}$, with a randomly chosen ungauged neighbour site.
\STATE Compute the fitness of the solutions obtained by crossover and mutation. Include these solutions in the current population, creating an augmented population.
\STATE \textit{Selection}: the population of solutions of the new generation will be selected from this augmented population. A proportion of solutions with best fitness, called elite, enter directly in the new generation while the remaining members of the new generation are randomly chosen according to certain fitness--dependent selection scheme (see Goldberg and Deb~\cite{GoldbergDeb} for a discussion of various selection schemes).
\STATE Stop the algorithm if the stop criterion is met. Otherwise, return to Step 3.
\end{algorithmic}
\end{algorithm}


\section{Determinantal Point Processes for Approximating The Optimum}\label{dpp}

Determinantal point processes are probabilistic models that capture negative correlation with respect to a similarity measure and offer efficient and exact algorithms for sampling, marginalization, conditioning, and other inference tasks. These process were first studied by Macchi \cite{Macchi}, as fermion processes, to model the distribution of fermions at thermal equilibrium. Borodin and Olshanski \cite{Borodin2000} as well as Hough et al. \cite{Hough} popularized the name ``determinantal" and gave probabilistic descriptions of DPPs. More recently, DPPs have attracted attention in the machine learning and statistics communities. The work of Kulesza and Taskar \cite{Kulesza and Taskar} provides a thorough and comprehensive introduction to the applications of DPPs that are most relevant to the machine learning community, such as image classification and document summarization.

\subsection{Definitions}
Recall that a point process $\mathbb{P}$ on the ground set $G = \{1, 2, ..., n\}$ is a probability measure defined on the power set of $G$, i.e., $2^{G}$. A point process $\mathbb{P}$ is called a determinantal point process, if when $Y$ is a random subset drawn according to $\mathbb{P}$, then we have for every $S \subseteq Y$,
\begin{equation}\label{marginal}
	\mathbb{P}(S \subseteq Y) = \text{det}(\matr{K}[S]),
\end{equation} for some matrix $\matr{K} \in \mathbb{R}^{n \times n}$
indexed by the elements of $G$ that is symmetric, real and positive semidefinite, and satisfies $0 \leq \bm{a}^T \matr{K} \bm{a} \leq 1$ for any $\bm{a} \in \mathbb{R}^{n \times 1}$.

In practice, it is more convenient to characterize DPPs via $L$--ensembles \cite{Borodin2005,Kulesza and Taskar}, which directly define the probability of observing each subset of $G$. An L--ensemble defines a DPP through a real positive semidefinite matrix $\matr{L}$, indexed by the elements of $G$, such that:
\begin{equation}\label{direct}
	\mathbb{P}_{\matr{L}}(\mathbf{Y} = Y) = \frac{\text{det}(\matr{L}[Y])}{\sum_{Y^\prime \subseteq G} \text{det}(\matr{L}[Y^\prime])},
\end{equation} where the normalizing constant $\sum_{Y^\prime \subseteq G} \text{det}(\matr{L}[Y^\prime]) = \text{det}(\matr{L} + \matr{I})$ and $I$ is an $n \times n$ identity matrix. Equation \eqref{direct} represents the probability of exactly observing all possible realizations of $\mathbf{Y}$. 

\subsection{$k$--Determinantal point processes}
Standard DPP models described above may yield subsets of any random size. A $k$--DPP on a discrete set $G = \{1, ..., n\}$ is simply a DPP with fixed cardinality $k$. It can be obtained by conditioning a standard DPP on the event that the set $Y$ has cardinality $k$, as follows
\begin{equation}\label{kdpp}
	\mathbb{P}_{\matr{L}}^k (Y) = \mathbb{P}(\mathbf{Y}=Y | |Y|=k) = \frac{\text{det}(\matr{L}[Y])}{\sum_{|Y^{\prime}|=k} \text{det}(\matr{L}[Y^{\prime}])},
\end{equation} where $|Y|$ denotes the cardinality of $Y$. This notion is essential in the context of our cardinality--constrained discrete optimization problem. We will show in the next subsection how we can sample from this probabilistic model and approach the optimal solution based on the sampling results.

\subsection{Sampling based solution strategy}
The sampling of a $k$--DPP largely relies on being able to express DPP as a mixture of elementary DPPs \cite{Kulesza and Taskar}, also commonly known as determinantal projection processes. Using Algorithm \ref{alg:kDPPsample} as adapted from Kulesza and Tasker \cite{Kulesza and Taskar}, the sampling from a $k$--DPP can be performed in $\mathcal{O}(N^3)$ time in general, and every $k$--element subset $S$ among the $n$ candidate points has the opportunity to be sampled with probability given in Equation \eqref{kdpp}.

To handle the NP--hard optimization problem in Equation \eqref{P}, the $k$--DPP sampling approach involves generating such $k$--DPP subsets repeatedly and calculating the objective function $v_{N} (S)$, such that successively better approximations, as measured by $v_{N} (S)$, can be found. The approximate solution to Problem \ref{P} is then given by the best $v_{N} (S)$ attained up to a certain number of simulations and its associated indices of points, as described in Algorithm \ref{alg:SamplingSoln}. 

Note that eigendecomposition of the kernel matrix can be done as a pre--processing step and therefore does not need to be performed before each sampling step. Therefore, assuming that we have an eigendecomposition of the kernel in advance, sampling one $k$--DPP run in $\mathcal{O}(Nk^3)$ time \cite{Kulesza and Taskar}, and the computation of the determinant of a submatrix typically takes $\mathcal{O}(k^3)$ time. Overall, Algorithm \ref{alg:SamplingSoln} runs in $\mathcal{O}(Nk^3)$ time per iteration.  

\begin{algorithm}[tbh]
\caption{Sampling from a $k$--DPP}
\label{alg:kDPPsample}
\begin{algorithmic}
\REQUIRE size $k$ and $\{{\bf v}_n, \lambda_n\}$ eigenvectors and eigenvalues of $\matr{L}$.
\STATE $J \leftarrow \emptyset$.
\STATE Compute elementary DPPs $E_{1}^{n}, \dots, E_{k}^{n}$, for $n = 0, \dots, N$.
\FOR{$n = N, \dots, 1$}
        \STATE Sample $u \sim U[0,1]$
        \IF{ $u < \frac{\lambda_n E_{k-1}^{n-1}}{E_{k}^{n}}$}
        \STATE $J \leftarrow J \cup \{n\}$
        \STATE $k \leftarrow k-1$
        \IF{k = 0}
        \STATE {\bf break}
        \ENDIF
        \ENDIF
\ENDFOR
\STATE $V \leftarrow \{{\bf v}_n\}_{n \in J}$
\STATE $Y \leftarrow \emptyset$
\WHILE{$|V| > 0$}
\STATE Select $y_i$ from $Y$ with probability given by $\frac{1}{|V|} \sum_{{\bf v} \in V} ({\bf v}^{\top} {\bf e}_i)^2$
\STATE $Y \leftarrow Y \cup \{y_i\}$
\STATE $V \leftarrow V_{\bot}$, an orthonormal basis for the subspace of $V$ orthogonal to ${\bf e}_i$
\ENDWHILE
\ENSURE $Y$.
\end{algorithmic}
\end{algorithm}

\begin{algorithm}[tbh]
\caption{Sampling--based solution strategy using $k$--DPP}
\label{alg:SamplingSoln}
\begin{algorithmic}[1]
\REQUIRE Size $k$ and the kernel matrix $\matr{L}$.
\STATE Sample $k$ indices according to the $k$-DPP distribution specified by $\matr{L}$ using Algorithm \ref{alg:kDPPsample}.
\STATE Compute determinant of the submatrix indexed by the $k$ indices sampled.
\STATE Repeat Step $1$ and $2$ until the maximum number of iterations or the maximum computing resources.
\ENSURE The maximum determinant and the associated set of indices.
\end{algorithmic}
\end{algorithm}

\section{Computational Performances}\label{examples}

In this section, we compare the performances of the greedy algorithm, the GA, and the $k$--DPP approach discussed above in three examples---a classical statistical design problem, a design of temperature monitoring network problem, and a large/intractable problem.

\subsection{D--optimal designs of experiments}
The statistical design problem amounts to selecting points in the multidimensional region that
will ``best'' estimate some important function of the parameters (see, e.g., Atkinson and Donev~\cite{Atkinson} or Federov~\cite{Federov} for a discussion of this topic). One of the most generally used is the D--criterion, which maximizes the determinant of $\matr{X}^T\matr{X}$ for a fixed number of design points, where $\matr{X}$ is the usual design matrix. Mathematically, suppose we have candidate points $\matr{X} \in \mathbb{R}^{n \times p}$, and the goal is to select a set of design points $S \subset N$ with $p \leq |S|=s \leq n$ such that the selected design satisfies the D--criterion. Using notation introduced in Equation~\eqref{P}, we have $v_N(S) = \text{det}(\matr{X}[S]\matr{X}[S]^T)$. 

\begin{figure}
    \centering
    \subfloat{
        \includegraphics[width=0.7\textwidth]{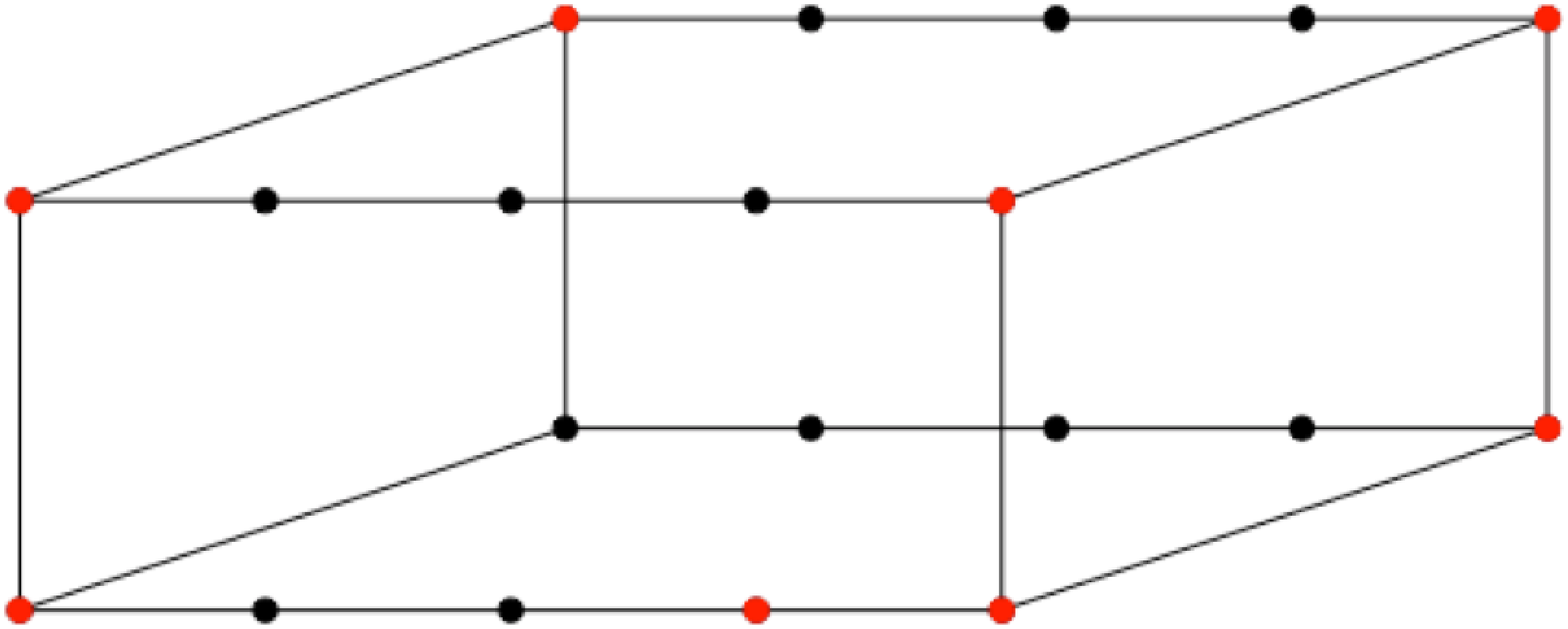}}
    \quad
    \subfloat{
        \includegraphics[width=0.7\textwidth]{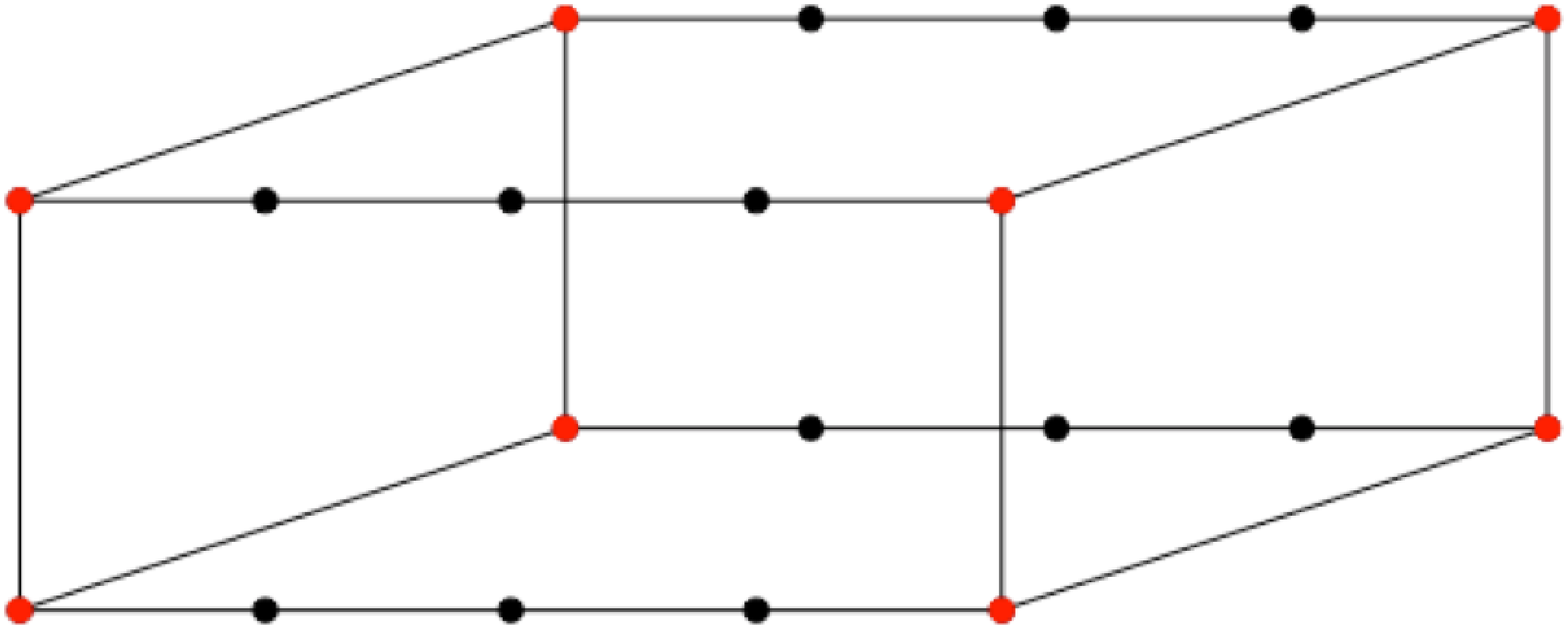}}
    \caption{Designs constructed by the greedy algorithm (top) and the $8$--DPP/GA (bottom). Solid dots are candidate points amongst which red solid dots are selected design points.}
    \label{fig:Ddesigns}
\end{figure}

Consider a simple model structure $\mathbb{E}\matr{Y} = \matr{X}\beta$ involving $3$ (factors) covariates with $5$, $2$, and $2$ levels, respectively. For our problem, the candidate set is a full factorial in all factors containing $20$ possible design points, and we select $8$ from them to form our design. Applying the greedy algorithm described in Ko et al.~\cite{KO1995}, the GA with tuning parameters $N_0 = 100$, $p_{\text{cross}} = 0.5$, $p_{\text{mut}} = 0.01$, and a tournament selection scheme with $4$ competitors for $1000$ generations, and an $8$--DPP for $10000$ iterations yield log-determinant of $7.407318$, $7.624619$, and $7.624619$, respectively. Note that the GA and the $8$--DPP both achieved the global optimum in this example. Figure~\ref{fig:Ddesigns} illustrates the constructed designs---the exact D-optimum design points fall on the vertices of the cube that spans the design space.

\subsection{Optimizing entropy based designs for monitoring networks}
For the first study we consider the data supplied by the U.S. Global Historical Climatology Network--Daily (GHCND), which is an integrated database of climate summaries from land surface stations across the globe. For illustrative purpose, we selected $97$ temperature monitoring stations where the maximum daily temperature was recorded. A subset of 67 stations were selected among the 97 stations to constitute a hypothetical monitoring network. An additional 30 stations were selected and designated as potential sites for new monitors. In this case study, the goal is to select a subset of 10 stations from among the 30 to augment the network based on the maximum entropy design criterion \cite{CaseltonZidek}. 

Using the notation in Equation~\eqref{P}, $K$ here is the estimated covariance matrix of $30$ candidate sites and $S$ is the subset of $10$ sites that maximize $v_N(S)$. For tractable optimization problems, the maximal value of the objective function (or equivalently the optimal design) can be obtained by exhaustive search. In this study, the maximal value is $80.09011$.

\begin{figure}
    \centering
	\includegraphics[width=\textwidth]{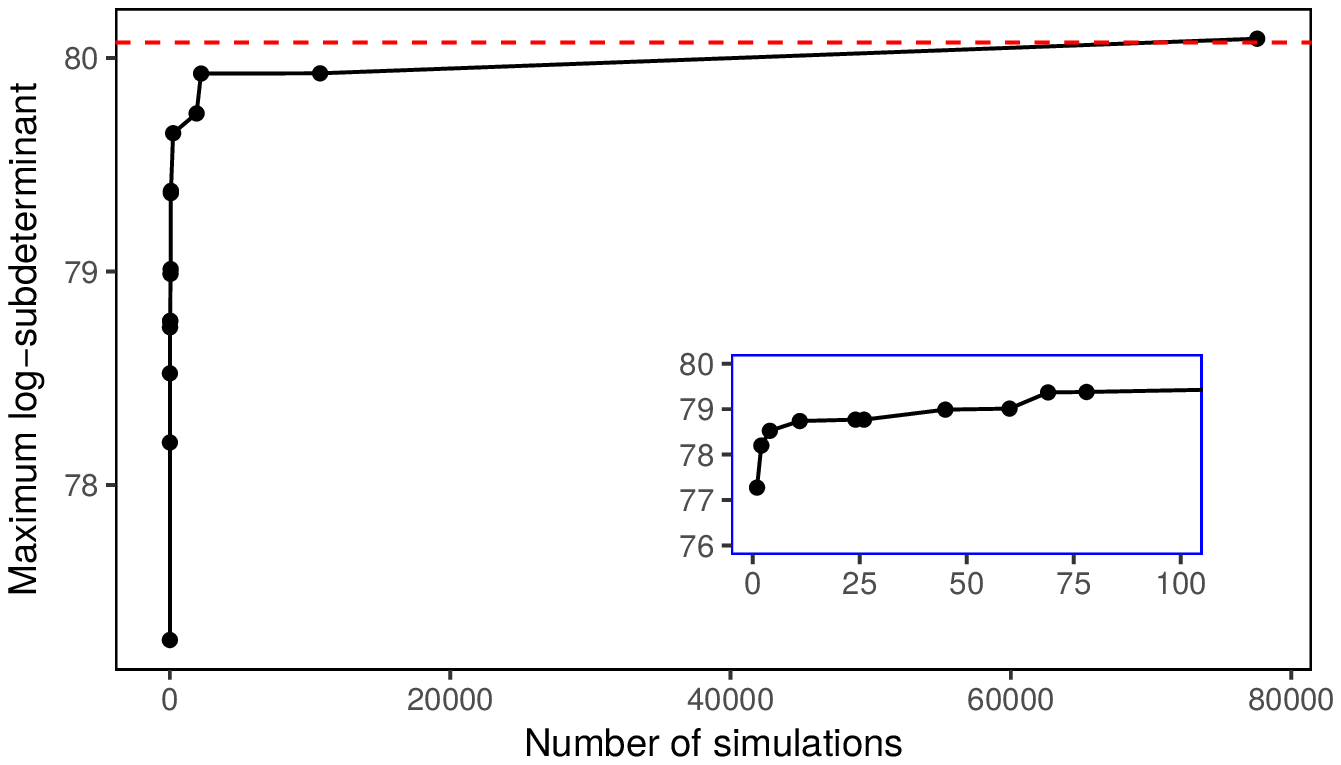}
	\caption{Occurrence of maximum log--determinants of the restricted conditional hypercovariance matrix when increasing the number of simulated $10$--DPP samples. The optimum solution is marked by the red horizontal dashed line. The inset is a zoomed--in version for the first $100$ samples.}
    \label{fig:dppreal}
\end{figure}

For the comparison, we first performed the greedy algorithm discussed in Jo et al.~\cite{KO1995}, which yielded a solution of $80.07284$. Using the tuning parameters suggested in Ruiz--C\'{a}rdenas et al.~\cite{Schmidt} which dealt with a similar design of monitoring network stations problem ($N_0 = 100$, $p_{\text{cross}} = 0.75$, $p_{\text{mut}} = 0.05$, and a tournament selection scheme with $4$ competitors), the GA yielded a solution of $80.09011$ after $1000$ generations. Similarly, the proposed $10$--DPP achieved the optimal value after about $80,000$ simulations. As illustrated in Figure~\ref{fig:dppreal}, the maximal value of the log--determinant among the simulations increases as the number of simulations gets larger.  

In terms of computation time, for this particular problem, it took about $20$ minutes of wall clock time to simulate $100,000$ subsets from the $10$--DPP using the \texttt{R} programming language \cite{R} on a laptop with a \texttt{2.5 GHz Intel Core i7} processor and a \texttt{16 GB 1600 MHz DDR3} RAM. In the same computational environment, it took $5$ minutes of wall clock time to simulate $1000$ generations of GA. 

\subsection{Synthetic data with a large number of points}
Exact methods quickly get inpractical for large data sets, and one has to resort to heuristics. Greedy heuristics are known to be fast and efficient, but they can be quite inaccurate. As an illustrative example, let an $n \times n$ real symmetric positive definite matrix be
\begin{equation*}
	A = 
	\begin{bmatrix}

    x_{11} & x_{12} & x_{13} & \dots  & x_{1n} \\
    x_{21} & x_{22} & x_{23} & \dots  & x_{2n} \\
    x_{31} & x_{32} & x_{33} & \dots  & x_{3n} \\
    \vdots & \vdots & \vdots & \ddots & \vdots \\
    x_{n1} & x_{n2} & x_{n3} & \dots  & x_{nn}

	\end{bmatrix},
\end{equation*} where the diagonal elements are
\begin{equation*}
	x_{ii} = 
	\begin{cases}
       d, & \text{if} \ i < n-k+1, \\
       d + \delta, & \text{otherwise.} \\
    \end{cases}
\end{equation*} The off-diagonal elements are
\begin{equation*}
	x_{ij} = 
	\begin{cases}
		a, & \text{if} \ i > j, i \neq n-k+1,n-k+2,...,n-k+10 \\
		b, & \text{if} \ i > j, i = n-k+1, \\
		c, & \text{if} \ i > j, i = n-k+2, \ \text{or} \ n-k+3, \ \text{or}..., n-k+10\\
		a, & \text{if} \ i < j, j \neq n-k+1,n-k+2,...,n-k+10 \\
		b, & \text{if} \ i < j, j = n-k+1, \\
		c, & \text{if} \ i < j, j = n-k+2, \ \text{or} \ n-k+3, \ \text{or}..., n-k+10,\\
	\end{cases}
\end{equation*} where $n$ is the size of the matrix and $k$ is the desired size of the subset one would like to select. 

\begin{figure}
    \centering
	\includegraphics[width=0.8\textwidth]{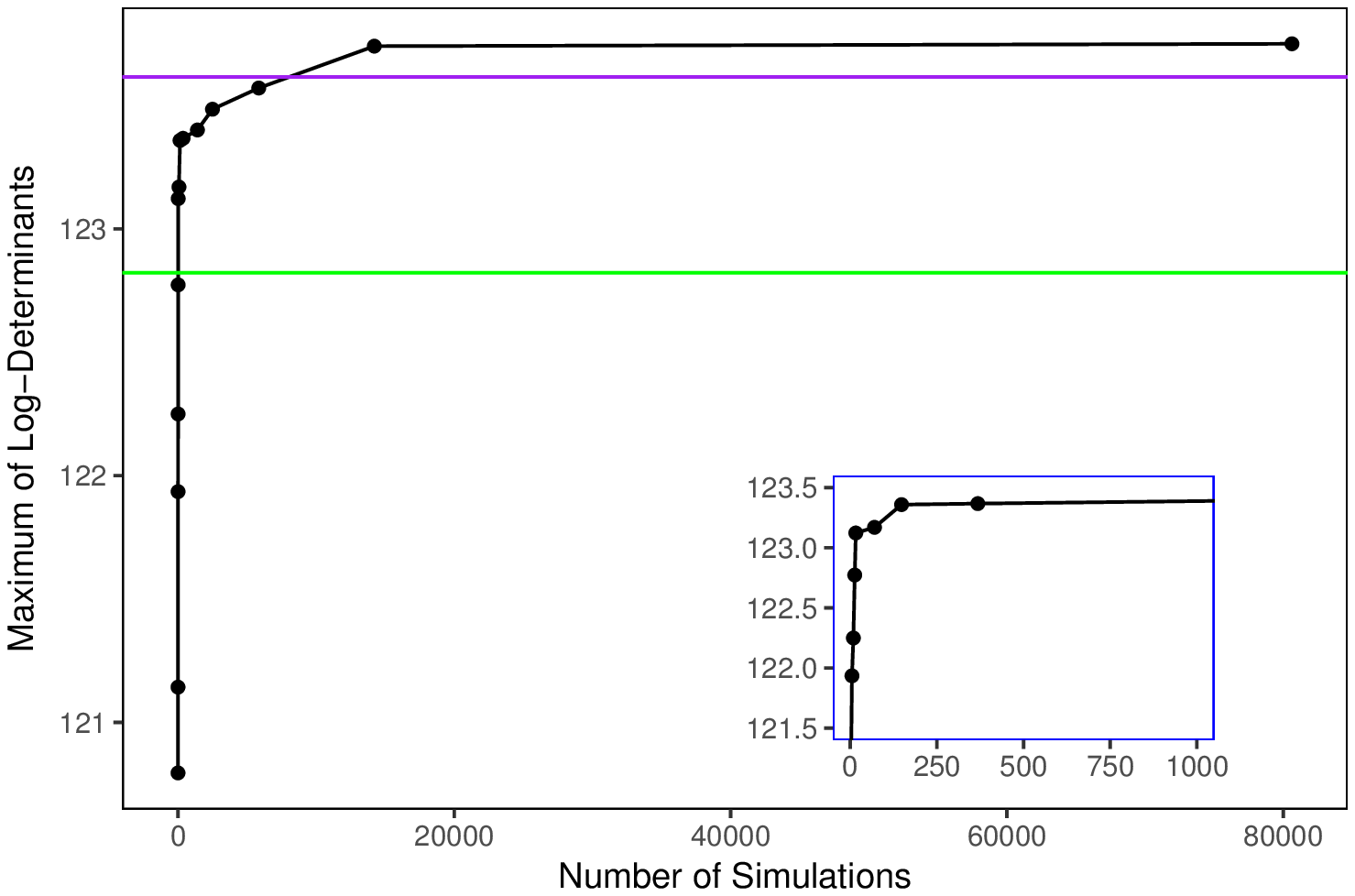}
	\caption{Occurrence of maximum log--subdeterminants of the synthetic kernel matrix when increasing the number of simulated $60$--DPP samples. The greedy and the GA solutions are marked by the green and purple horizontal solid lines, respectively. The inset is a zoomed--in version for the first $1000$ samples.}
    \label{fig:dppsyn}
\end{figure}

Suppose we seek a $60$--by--$60$ submatrix with maximal determinant from a $100$--by--$100$ matrix, with $a=0.2$, $b=0.9$, $c=0.65$, $d=7$ and $\delta=1$. For this particular matrix, running a greedy algorithm results in the selection of subsets $\{31,...,40,51,...,100\}$ at termination, and an associated log--subdeterminant of $122.8217$. We also ran GA with the same tuning parameters as in the previous section and the best solution obtained was $123.6158$ in $1000$ iterations. For comparison, we simulated $100,000$ $60$--DPP samples and a better solution of $123.7503$ is found, as shown in Figure \ref{fig:dppsyn}.

The computational burden increases significantly when dealing with large matrices, but parallel simulations can be exploited to reduce the computational time. For this example, it took $1$ hour of wall clock time to simulate $100,000$ samples of $60$--DPP on a Compute Canada cluster with \texttt{32 cores 2.1GHz Intel Broadwell} CPUs and \texttt{128 GB} RAM. In the same computational environment, running $1,000$ generations of GA required $1$ hour of wall clock time. 

In summary, the GA and the DPP approaches gave fairly comparable solutions that are better than those produced by the greedy algorithm. The DPP methods seemed to require more computing resources. To see the variations from run to run, we repeated the last two case studies $100$ times with the GA and the DPP approaches. The results for DPP with $100,000$ and $1,000,000$ simulation and the GA with $1,000$ and $10,000$ generations are shown in Table~\ref{tab:t1}. Overall, the performances of the two approaches are fairly comparable.   

\begin{table}[p]
    \centering
	\begin{tabular}{lcccccc}
	\toprule
	\multirow{2}[3]{*}{Sample Size} & \multicolumn{3}{c}{$k=10$} & \multicolumn{3}{c}{$k=60$} \\
	\cmidrule(lrr){2-4} \cmidrule(lrr){5-7}
 	& Maximum & Mean & SD & Maximum & Mean & SD \\
	\midrule
	DPP-100,000 & 80.09 & 80.00 & 0.05 & 123.75 & 123.72 & 0.05 \\
	DPP-1,000,000 & 80.09 & 80.07 & 0.01 & 123.89 & 123.80 & 0.02 \\
	GA-1,000 & 80.09 & 80.07 & 0.06 & 123.62 & 123.40 & 0.15 \\
	GA-10,000 & 80.09 & 80.08 & 0.01 & 124.02 & 123.89 & 0.08 \\
	\bottomrule
	\end{tabular}
	\caption{Results obtained from $100$ realizations of $k$-DPP and GA for the design of monitoring networks example and the synthetic data example. Sample size refers to the number of DPP simulations and the number of generations of GA for each realization; SD refers to standard deviation.}\label{tab:t1}
\end{table}

\section{Discussion}
This paper introduces a sampling based approach for approximating the combinatorial optimization problem of subdeterminant maximization. By sampling from a $k$--DPP, which can be done in polynomial time, we approach the optimal solution by using the maximum  simulated value as an approximation.

We demonstrated the potential applications of the $k$--DPP based algorithm for constructing optimal designs of experiment and finding optimal allocation of spatial monitoring networks, and found it successful in obtaining the exact solution for small, tractable problems. When the size of the problem makes exact methods impractical, we showed (for a certain type of matrix) that our algorithm outperforms the greedy algorithm and is comparable to the genetic algorithm for a relatively small cost in computational time. When solving a large problem where the exact methods do not work, the proposed DPP method is guaranteed to ultimately approach the optimum with a sufficient number of iterations while the greedy and GA potentially converge to a local optimum after a certain number of iterations. Moreover, although GA usually runs faster in computational time than our DPP approximation and gives fairly accurate solutions, it requires careful calibrations of its tuning parameters. In fact, finding the right balance between crossover/selection, which pulls the population towards a local maximum, and mutation, which explore potentially better solution spaces, is a known issue for GA---inappropriate choices of tuning parameters could adversely affect the convergence of the algorithm, see Goldberg and Holland~\cite{Goldberg} and Whitley~\cite{Whitley} for detailed discussions. The DPP approximation, on the other hand, can be run naively to obtain comparably accurate solutions. Another major advantage of the DPP is that the algorithm can be easily implemented with parallelization without material modifications, which provides potential usage of free--access supercomputers to further reduce computational time.

In future work, approximate sampling algorithms for $k$--DPP will be explored. Work has been published recently trying to reduce the dimension of the matrix, such as the one introduced in Li, Jegelka, and Sra \cite{Li}. Others mainly focus on the approximations of the kernel matrix using some lower dimensional structures or alternate representations of the matrix in lower dimensional forms, such as the ones in Affandi et al. \cite{Affandi1} and Kulesza and Taskar \cite{Kulesza}. These methods would help reduce the sampling complexity of the $k$--DPPs and eventually could further reduce the computational time in obtaining the approximate solutions. In current work analytical theory is being developed to describe the number of iterations needed for successive improvements in the approximate DPP solutions as well as to estimate the expected duration of time until an optimal solution is obtained.


\end{document}